# Revealing the conduction band and pseudovector potential in 2D moiré semiconductors

Abigail J. Graham[1], Heonjoon Park[2], Paul V. Nguyen[2], James Nunn[1], Viktor Kandyba[3], Mattia Cattelan[3], Alessio Giampietri[3], Alexei Barinov[3], Kenji Watanabe[4], Takashi Taniguchi[5], Anton Andreev[2], Mark Rudner[2], Xiaodong Xu[2,6], Neil R. Wilson[1], & David H. Cobden[2]

[1] Department of Physics, University of Warwick, Coventry, CV4 7AL, U.K.
[2] Department of Physics, University of Washington, Seattle, WA, USA.
[3] Elettra – Sincrotrone Trieste, S.C.p.A, Basovizza (TS), 34149, Italy.
[4] Research Center for Functional Materials, National Institute for Materials Science, 1-1 Namiki, Tsukuba 305-0044, Japan.
[5] International Center for Materials Nanoarchitectronics, National Institute for Materials Science, 1-1 Naniki, Tsukuba 305-0044, Japan.
[6] Department of Materials Science and Engineering, University of Washington, Seattle, WA, USA.

**Abstract**

Stacking monolayer semiconductors results in moiré patterns that host many correlated and topological electronic phenomena, but measurements of the basic electronic structure underpinning these phenomena are scarce. Here, we investigate the properties of the conduction band in moiré heterobilayers using submicron angle-resolved photoemission spectroscopy with electrostatic gating, focusing on the example of $WS_2/WSe_2$. We find that at all twist angles the conduction band edge is the K-point valley of the $WS_2$, with a band gap of $1.58 \pm 0.03$ eV. By resolving the conduction band dispersion, we observe an unexpectedly small effective mass of $0.15 \pm 0.02\ m_e$. In addition, we observe replicas of the conduction band displaced by reciprocal lattice vectors of the moiré superlattice. We present arguments and evidence that the replicas are due to modification of the conduction band states by the moiré potential rather than to final-state diffraction. Interestingly, the replicas display an intensity pattern with reduced, 3-fold symmetry, which we show implicates the pseudo vector potential associated with in-plane strain in moiré band formation.

**Introduction**

The diverse ramifications of moiré superstructures in two-dimensional (2D) van der Waals heterostructures are of great current interest. Most famously, stacks of graphene sheets with appropriate rotational misalignment between the layers exhibit moiré superlattices that create nearly flat bands and lead to correlated insulating states, superconductivity, Chern insulators, and more[1–4]. The existence of these graphene moiré bands, and of correlation-induced spectral gaps within them, has been directly confirmed by submicron-scale angle-resolved photoemission spectroscopy[5,6] (μARPES) and scanning tunneling microscopy[7,8].

Artificial bilayers of two-dimensional (2D) semiconductors also exhibit moiré superlattices[9,10] leading to exciton arrays[11–14], Mott insulating states and generalized Wigner crystals[15,16], excitonic insulators[17,18], tuneable magnetism[16,19,20], Kondo lattices[21], and very recently fractional quantum anomalous Hall states[22–24]. In the present work we use μARPES to probe the band structure of such 2D moiré semiconductors. Although the conduction bands (CBs) play a crucial role in many of the above phenomena, ARPES detects only occupied states and thus is normally limited to probing the valence bands[25–27]. To overcome this limitation, we incorporate a metallic gate electrode under the heterostructure, which allows electrostatic doping and thus detection of the CB edges[28] as well as changes in the bands resulting from doping[29–31] or electric field[32].

We focus on WS$_2$/WSe$_2$ heterobilayers, where the moiré potential is known to be strong at small twist angles[16,33]. We determine the fundamental conduction band parameters, show that the band alignment of the separate monolayers is maintained independent of twist angle, determine the band gap and the CB effective mass, and observe perturbing effects of the moiré potential on the CB. The latter manifest as multiple replicas of the original CB displaced by reciprocal lattice vectors of the moiré superlattice. We consider the expected relative contributions of such moiré potential-induced reconstruction of the CB states and of "final-state diffraction" of photoemitted electrons by the moiré potential as they exit the material, concluding that the CB state reconstruction effect should be dominant at small twist angles. We notice that the replicas display a pronounced alternation of intensity when tracing them around the K-point, exhibiting only 3-fold rotational symmetry rather than the 6-fold symmetry one would expect for scattering of a free particle off a triangular or honeycomb lattice. For a model starting with circularly symmetric dispersion, as appropriate for low energies near the bottom of the WS$_2$ conduction band, we show that such an intensity pattern with reduced rotation symmetry cannot be produced within a model that incorporates a purely scalar moiré potential. Our results thus reveal a significant influence of the moiré pseudovector potential that is expected to be present as a result of strain.

*Samples and measurements.* Devices were fabricated by mechanical exfoliation, dry transfer, and electron-beam patterning of metal electrodes as in previous work[28]. A graphene top contact, pre-shaped into a comb-like pattern using atomic force microscope (AFM)-based electrochemical patterning[34], overlaps the heterobilayer which lies on a thin flake of insulating hexagonal boron nitride (hBN) over a graphite back gate, supported in turn on a SiO$_2$/Si chip (see Fig. 1a and Methods). Each semiconductor heterobilayer was constructed by placing monolayer flakes with their straight edges subtending a target angle. The actual angles obtained were determined from the μARPES spectra via identification of the constituent layers' valence band edges at their zone corners, but could be inferred from the moiré period revealed by piezo-force microscopy[35] (PFM; see SI Sec. 1).

The lattice constants of relaxed monolayer WS$_2$ and WSe$_2$ are $a_{WS_2} = 0.315$ nm and $a_{WSe_2} = 0.328$ nm respectively, and the lattice mismatch parameter is $\delta = (a_{WSe_2}/a_{WSe_2}) - 1 = 0.041$. The moiré lattice constant[36], $a_m = a_{WSe_2}[2(1+\delta)(1-\cos\phi) + \delta^2]^{-1/2}$, has its largest value of $a_{WSe_2}/\delta \approx 8$ nm [16] at $\phi = 0$. On device 1, photoluminescence measurements (SI Sec. 2) show enhancements in intensity at gate voltages corresponding to the integer filling of a moiré unit cell of area consistent with the twist angle independently inferred by PFM ($a_m \sim 2.8$ nm, $\phi = 6°$). Note, however, that we cannot tell whether the stacking is in the antiparallel (centrosymmetric) or parallel (polar) configuration from any of these measurements.

For μARPES measurements, a device would be mounted on the temperature stage at ~100 K with the top graphene connected to ground through a current amplifier and a voltage $V_g$ applied to the back gate. At high $V_g$ photoemission near the Fermi level $E_F$ from the semiconductors can only obtained when the submicron beam spot (27 eV photon energy) is focused between the teeth of the graphene comb, as illustrated in Fig. 1b (see SI Sec. 3 for details).

**Results and Discussion**

*Valence bands.* Figure 1c is a sketch of the Brillouin zones for the heterobilayer in device 1. Figures 1d and e show energy-momentum slices measured at $V_g = 0$ along the high symmetry directions $\mathbf{\Gamma} - \mathbf{K}_{WS_2}$ and $\mathbf{\Gamma} - \mathbf{K}_{WSe_2}$, respectively. As usual, we plot the energy relative to $E_F$, i.e., $E - E_F$ (Methods). The bands near the zone corner closely match the spin-split valence bands (VBs) of isolated WS$_2$ and WSe$_2$ monolayers (SI Sec. 4), implying weak hybridization far from $\mathbf{\Gamma}$, as expected. The overlaid dotted lines are fits to the upper WSe$_2$-like band (red) and the upper and lower WS$_2$-like bands (blue), yielding hole effective masses of $0.47 \pm .02\, m_e$, $0.38 \pm .01\, m_e$, and $0.56 \pm .01\, m_e$ respectively ($m_e$ is the free electron mass). The WS$_2$ spin-orbit splitting is $\Delta_{SO}^{WS_2}$ = 0.44 ± 0.04 eV, the same as in the monolayer[37–39]. The VB edge is the WSe$_2$-like band at $\mathbf{K}_{WSe_2}$, which is 0.58 ± 0.04 eV above the WS$_2$-like band maximum at $\mathbf{K}_{WS_2}$. These band parameters do not vary noticeably with twist angle (SI Sec. 5). When the WS$_2$ is on top, the WS$_2$-like bands near the zone boundary are more intense; this is explained by weak interlayer hybridization and the rapid fall-off of photoemission strength with depth. Indeed, when the WSe$_2$ is on top the converse is seen (see SI Sec. 6). In contrast, near $\mathbf{\Gamma}$ two bands with similar intensity are seen; this is explained by strong interlayer hybridization at $\mathbf{\Gamma}$.[25]

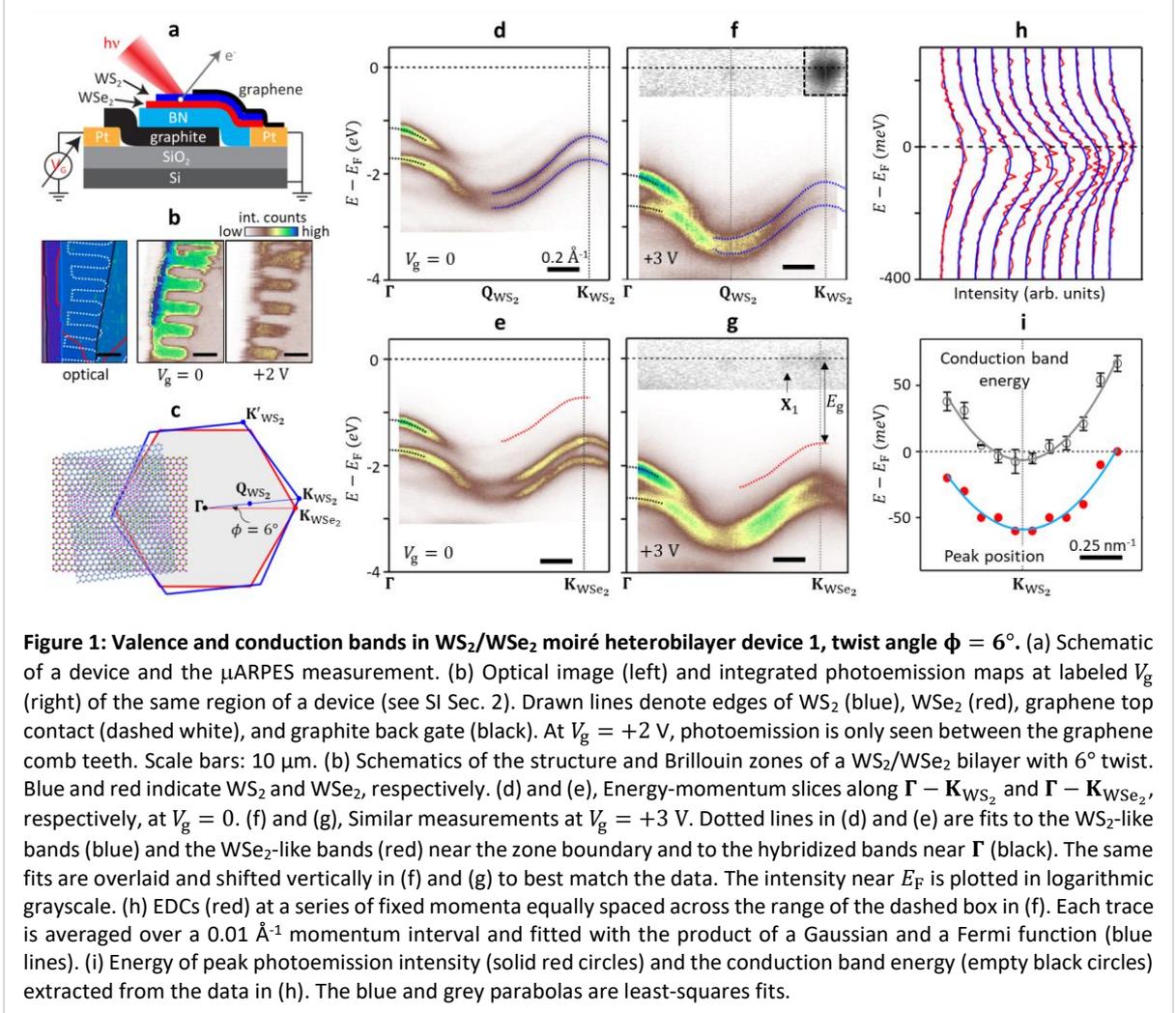

**Figure 1: Valence and conduction bands in WS$_2$/WSe$_2$ moiré heterobilayer device 1, twist angle $\phi = 6°$.** (a) Schematic of a device and the μARPS measurement. (b) Optical image (left) and integrated photoemission maps at labeled $V_g$ (right) of the same region of a device (see SI Sec. 2). Drawn lines denote edges of WS$_2$ (blue), WSe$_2$ (red), graphene top contact (dashed white), and graphite back gate (black). At $V_g = +2$ V, photoemission is only seen between the graphene comb teeth. Scale bars: 10 μm. (b) Schematics of the structure and Brillouin zones of a WS$_2$/WSe$_2$ bilayer with 6° twist. Blue and red indicate WS$_2$ and WSe$_2$, respectively. (d) and (e), Energy-momentum slices along $\mathbf{\Gamma} - \mathbf{K}_{WS_2}$ and $\mathbf{\Gamma} - \mathbf{K}_{WSe_2}$, respectively, at $V_g = 0$. (f) and (g), Similar measurements at $V_g = +3$ V. Dotted lines in (d) and (e) are fits to the WS$_2$-like bands (blue) and the WSe$_2$-like bands (red) near the zone boundary and to the hybridized bands near $\mathbf{\Gamma}$ (black). The same fits are overlaid and shifted vertically in (f) and (g) to best match the data. The intensity near $E_F$ is plotted in logarithmic grayscale. (h) EDCs (red) at a series of fixed momenta equally spaced across the range of the dashed box in (f). Each trace is averaged over a 0.01 Å$^{-1}$ momentum interval and fitted with the product of a Gaussian and a Fermi function (blue lines). (i) Energy of peak photoemission intensity (solid red circles) and the conduction band energy (empty black circles) extracted from the data in (h). The blue and grey parabolas are least-squares fits.

*Band gap.* Figs. 1f and g show corresponding measurements made at a positive voltage $V_g = +3$ V, which capacitively induces electron doping $n_g = (6.4 \pm 0.4) \times 10^{12}$ cm$^{-2}$ (Methods). Photoemission can now be seen from the CB edge near $E_F$. Note that there is a broadening of all features relative to the $V_g = 0$ data which can be explained by the varying electrostatic potential over the beam spot associated with the in-plane current flow that is required to replenish the photoemitted charge. Strong CB emission is seen at $\mathbf{K}_{WS_2}$ in Fig. 1d, while much weaker emission is seen at $\mathbf{Q}_{WS_2}$ implying that the CB minimum at $\mathbf{Q}_{WS_2}$ is close to but higher than the one at $\mathbf{K}_{WS_2}$ (by ~10-20 meV; see Methods). The absolute band gap at this doping is $E_g$ = 1.58 ±0.03 eV, while the intralayer gap between the WS$_2$-like bands at $\mathbf{K}_{WS_2}$ is 2.04 ±0.03 eV, consistent with the gap of monolayer WS$_2$ measured at $n_g = (1.0 \pm 0.2) \times 10^{12}$ cm$^{-2}$ in prior work[28]. These CB parameters, like the VB ones mentioned above, did not vary detectably with twist angle (SI Sec. 5).

*Conduction bands.* The gate doping achieved in device 1 was sufficient to determine the CB curvature, which has not been done before. Fig. 1h shows energy dispersion curves (EDCs) through the CB feature in the dashed box in the top right corner of Fig. 1f. The energy where the intensity is maximum, plotted as solid red circles in Fig. 1h, passes through a minimum at $\mathbf{K}_{WS_2}$. The spin splitting of the CB is several times $k_B T$ at 100 K [37,40], so we assume the lower spin branch is mainly populated and derive its dispersion by fitting the EDC at each momentum to the product of a Fermi function ($T = 100$ K, $E_F = 0$) and a Gaussian (width 160 meV) treating the Gaussian center $E_c$ as a fitting parameter (see Methods). The resulting $E_c$ values are plotted as open circles in Fig. 1i. Fitting a parabola (black line) yields an effective mass $m_e^* = 0.15 \pm 0.02\, m_e$. We note that this is substantially smaller than first-principles predictions for monolayer $WS_2$ [37,41], which lie in the range $0.24 - 0.27\, m_e$.

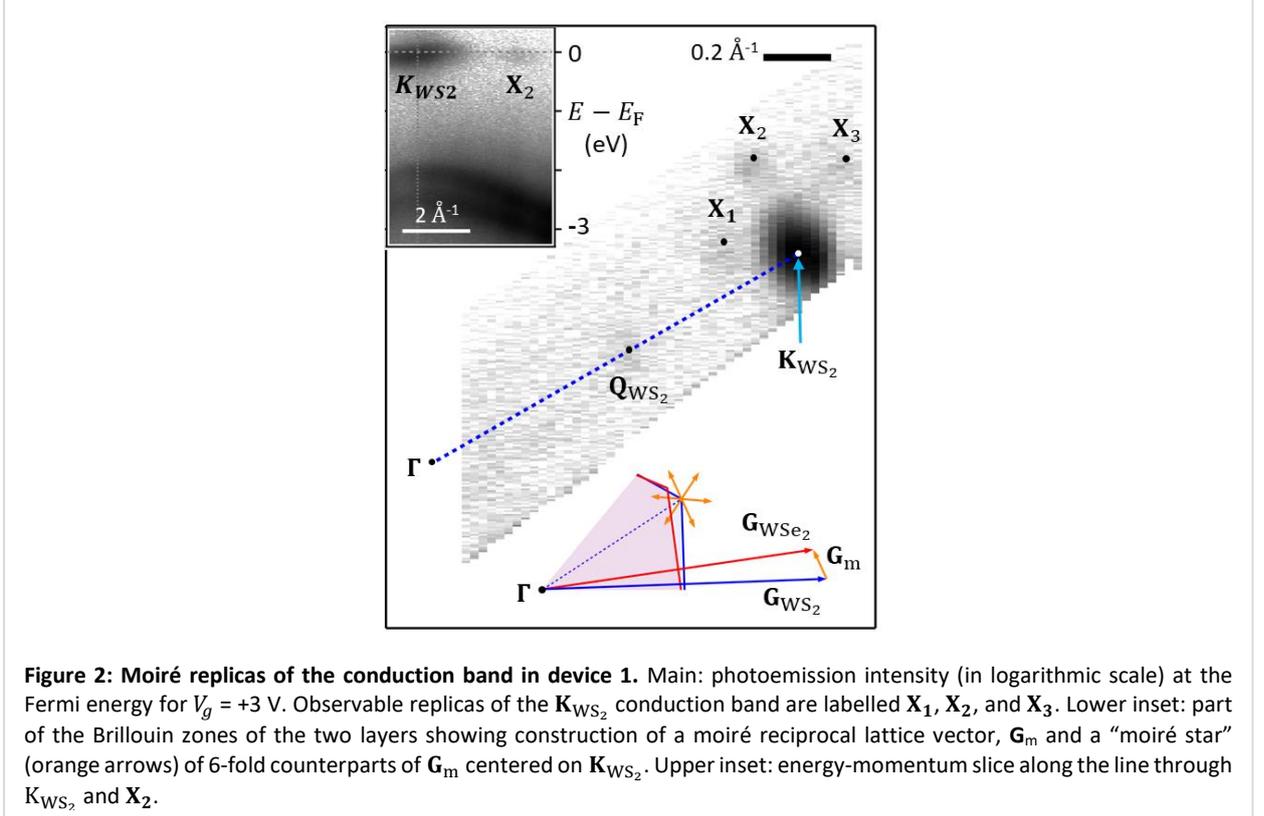

**Figure 2: Moiré replicas of the conduction band in device 1.** Main: photoemission intensity (in logarithmic scale) at the Fermi energy for $V_g$ = +3 V. Observable replicas of the $\mathbf{K}_{WS_2}$ conduction band are labelled $\mathbf{X_1}$, $\mathbf{X_2}$, and $\mathbf{X_3}$. Lower inset: part of the Brillouin zones of the two layers showing construction of a moiré reciprocal lattice vector, $\mathbf{G_m}$ and a "moiré star" (orange arrows) of 6-fold counterparts of $\mathbf{G_m}$ centered on $\mathbf{K}_{WS_2}$. Upper inset: energy-momentum slice along the line through $\mathbf{K}_{WS_2}$ and $\mathbf{X_2}$.

*Replicas.* In Fig. 1g, we discern an additional spot of emission near the Fermi energy, labeled $\mathbf{X_1}$, that does not correspond to the band edge of either constituent monolayer. Figure 2 is a constant-energy map at $E = E_F$ in which $\mathbf{X_1}$ is seen as one of three satellite spots situated near the corners of a hexagon centered on $\mathbf{K}_{WS_2}$. The two others are labelled $\mathbf{X_2}$ and $\mathbf{X_3}$. These spots appear at $E_F$ simultaneously with the CB minimum at $\mathbf{K}_{WS_2}$, as illustrated by the momentum slice passing through $\mathbf{K}_{WS_2}$ and $\mathbf{X_2}$ shown in the upper inset. To within uncertainty, they are displaced from $\mathbf{K}_{WS_2}$ by moiré reciprocal lattice vectors $\mathbf{G_m}$. The latter are determined by the relation $\mathbf{G_m} = \mathbf{G}_{WSe_2} - \mathbf{G}_{WS_2}$, where $G_{WSe_2}$ and $G_{WS_2}$ are reciprocal lattice vectors of the two layers, as illustrated in the lower inset. In this device, $G_m = 2\pi/a_m = 2.5$ nm$^{-1}$. The corresponding value of $a_m$ of 2.5 nm was confirmed by PFM imaging of the device. We deduce that the satellite spots are replicas of the CB minimum related to the moiré pattern. Similar moiré-related replicas of the VB have been reported in photoemission from WSe$_2$ under graphene[42] and WS$_2$ under graphene[43] and interpreted in terms of miniband formation.

Replicas of the CB were also seen in device 2 ($\phi \sim 2°$, $G_m = 0.9$ nm$^{-1}$). Fig. 3a is an energy-momentum slice from the heterobilayer in device 2 along $\mathbf{\Gamma} - \mathbf{K}_{WS_2}$ at $V_g = +2.5$ V ($n_g = (4.2 \pm 0.4) \times 10^{12}$ cm$^{-2}$), and Fig. 3b is a constant-energy map around $\mathbf{K}_{WS_2}$ at $E = E_F$. The CB feature here has three lobes that are consistent with partially resolved replicas of a central spot displaced by three moiré reciprocal lattice vectors, one of which is constructed Fig. 3c. Notably, replicas were also seen in the spectrum of graphene overlapping the heterobilayer. Its Brillouin zone (rotated by 19° relative to the WS$_2$) is also shown in Fig. 3c. Fig. 3d shows an energy-momentum slice through the graphene zone corner, $\mathbf{K}_g$, and Fig. 3e shows corresponding constant-energy maps at the indicated energies. In addition to the ordinary Dirac cone centered at $\mathbf{K}_g$ there is a set of replicas around it that form a slightly distorted triangular array. The same three moiré vectors match the heterobilayer CB replicas in Fig. 3b and the more intense graphene replicas in Fig. 3e, implying that all are related to the WS$_2$/WSe$_2$ moiré pattern. Similar patterns were seen before in ARPES measurements[44] on (ungated) graphene on WS$_2$/WSe$_2$, where the authors also pointed out that the distortion could be due to anisotropic strain. We saw a similar pattern again in measurements on graphene overlapping a WS$_2$/MoSe$_2$ heterobilayer (SI Sec. 7).

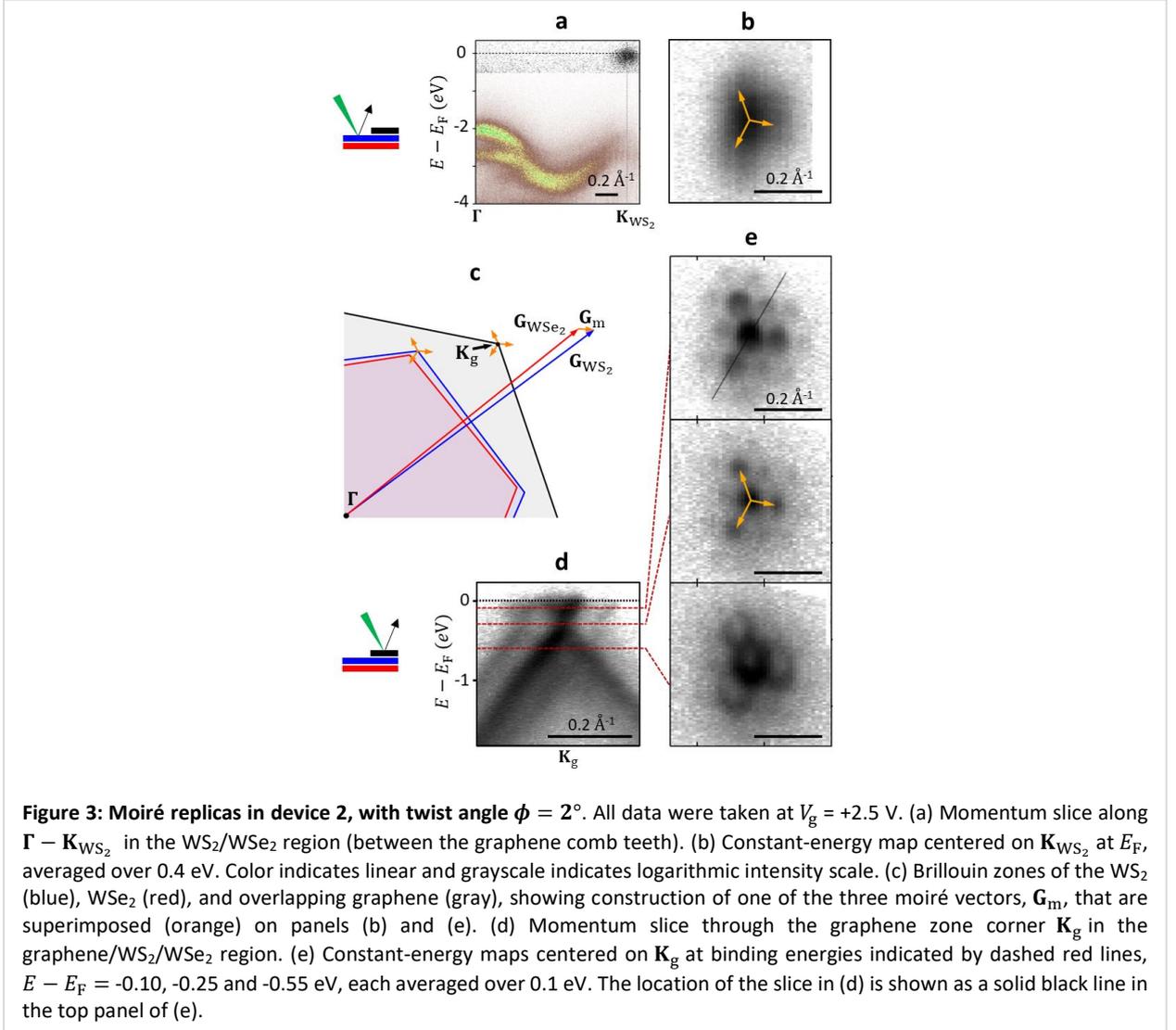

**Figure 3: Moiré replicas in device 2, with twist angle $\phi = 2°$.** All data were taken at $V_g$ = +2.5 V. (a) Momentum slice along $\mathbf{\Gamma} - \mathbf{K}_{\text{WS}_2}$ in the WS$_2$/WSe$_2$ region (between the graphene comb teeth). (b) Constant-energy map centered on $\mathbf{K}_{\text{WS}_2}$ at $E_F$, averaged over 0.4 eV. Color indicates linear and grayscale indicates logarithmic intensity scale. (c) Brillouin zones of the WS$_2$ (blue), WSe$_2$ (red), and overlapping graphene (gray), showing construction of one of the three moiré vectors, $\mathbf{G}_m$, that are superimposed (orange) on panels (b) and (e). (d) Momentum slice through the graphene zone corner $\mathbf{K}_g$ in the graphene/WS$_2$/WSe$_2$ region. (e) Constant-energy maps centered on $\mathbf{K}_g$ at binding energies indicated by dashed red lines, $E - E_F$ = -0.10, -0.25 and -0.55 eV, each averaged over 0.1 eV. The location of the slice in (d) is shown as a solid black line in the top panel of (e).

In higher-twist device 3 ($\phi = 9°$, $G_m = 3.6$ nm$^{-1}$; see SI Sec. 8) no CB replicas were visible. On the other hand, this device was the only one that exhibited VB replicas. This could be just a matter of energy resolution: for example, we estimate that for $\phi = 6°$ a resolution of ~100 meV is needed to distinguish VB replicas compared with ~400 meV for CB replicas, because of the smaller dispersion of the VB. Under no conditions did we see replicas associated with moiré wavevectors of the graphene/WS$_2$ interface where the large lattice mismatch should make moiré modulations very small. This argues in favor of a role for scattering from the moiré potential.

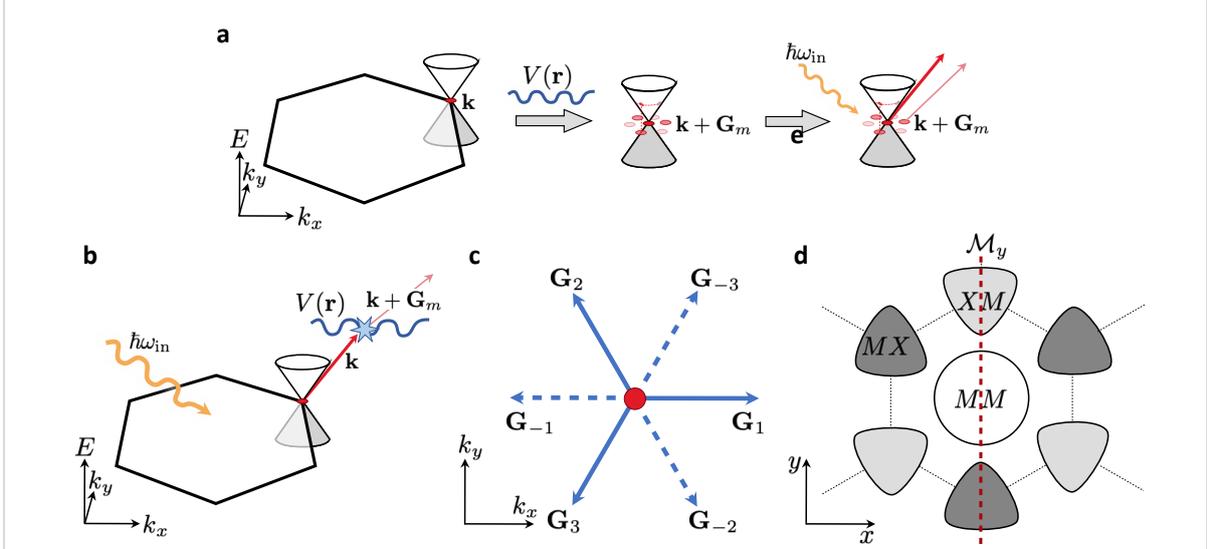

**Figure 4: Origin of the moiré replicas.** Illustrations with Dirac cones represent the behavior within the graphene layer on top of the WS$_2$/WSe$_2$ heterostructure, but the same mechanisms apply for the WS$_2$ layer. (a) Initial-state modification. Bloch states of the superlattice associated with the moiré potential V(**r**) are formed from superpositions of states in the unperturbed conduction band, offset by reciprocal lattice vectors of the moiré pattern, {**G**$_m$}. Photoemission from the superlattice Bloch states thus carries in-plane momentum contributions both from a central peak corresponding to the original conduction band and satellites (replicas) of weaker intensity that map out the momentum space structure of the reconstructed conduction band. (b) Final-state diffraction. Ignoring the effects of the moiré superlattice on the conduction band states themselves, the moiré potential may also scatter photoemitted electrons during their escape from the material, producing replica intensity spots displaced from the main peak by moiré reciprocal lattice vectors. As described in the text and SI Sec. 9, for small twist angles and high photoexcitation energies, we expect the observed replica intensity to be dominated by the initial state modification effect. (c) The six shortest reciprocal lattice vectors of the $C_3$-symmetric moiré pattern. (d) Schematic representation of the moiré unit cell of the WS$_2$/WSe$_2$ heterostructure with $C_{3v}$ symmetry. The shading indicates different values of the scalar moiré potential $U(\boldsymbol{r})$ near the high symmetry regions *MM* where the metal (W) atoms are vertically aligned, and the MX and XM regions where the metal atom of one layer sits directly above or below the chalcogen atom of the opposite layer.

*Origin of the CB replicas.* All of the replica features mentioned above appear to be copies of the parent bands translated by reciprocal lattice vectors of the moiré pattern of the heterobilayer. In general, these replicas result from the combination of moiré potential-induced modifications of the system's Bloch states ("initial-state modification" or "miniband formation"), as indicated schematically in Fig. 4a, and scattering of the photoexcited electrons by the moiré potential as they leave the sample ("final-state diffraction")[45–49], as indicated in Fig. 4b. We now briefly discuss the qualitative features of these two contributions, and the factors that point to initial-state modification as the dominant source of the replicas. Our discussion applies both to replicas seen in the CB of WS$_2$ and to those observed for the graphene on top of the WS$_2$/WSe$_2$ heterostructure as seen in Fig. 3.

Initial-state modification results from electrons coherently scattering on the moiré potential. New Bloch states of the superlattice are formed by hybridizing states in the original bands of the material at crystal momentum values offset by integer linear combinations of the moiré reciprocal lattice vectors {**G**$_m$}; see Fig. 4c (where we show the six shortest **G$_m$**). From perturbation theory, it is straightforward to see that this hybridization is strongest when the energy differences between states offset by a moiré wavevector are small (compared with the strength of the effective moiré potential, $|U|$). Thus, initial state modification is stronger when the moiré reciprocal lattice vectors are shorter, that is, for smaller twist angles. Indeed, the CB replicas are strongest in device 2 ($\phi = 2°$; Fig. 3), weaker in device 1 ($\phi = 6°$; Fig. 2), and not detectable in device 3 ($\phi = 9°$).

The magnitudes of the final-state diffraction contributions are determined by the corresponding differential cross-sections for the photo-emitted electrons to scatter from the moiré potential. Although the interaction between the photo-emitted electron and the material may be strong, due to the emitted electron's high velocity the interaction time is short. In terms of the moiré potential amplitude $U$ (see below for further microscopic details), the amplitude corresponding to the scattering process is controlled by the parameter $Ud/(\hbar v_{out})$, where $v_{out}$ is the velocity of the emitted electron and $d$ is the distance over which the moiré potential acts. For $Ud/(\hbar v_{out}) \ll 1$, the scattering amplitude may be estimated as $\mathcal{A}_{fin} \sim Ud/(\hbar v_{out})$. For comparison, consider an electronic state at momentum $\boldsymbol{k}$ (in the absence of the moiré potential); in the presence of the moiré potential, the wave function of this state obtains a component at momentum $\boldsymbol{k} + \mathbf{G}_{\mathrm{m}}$ that in the perturbative regime can be estimated as $\mathcal{A}_{ini} \sim \frac{U}{[\varepsilon(\boldsymbol{k}) - \varepsilon(\boldsymbol{k}+\mathbf{G}_{\mathrm{m}})]}$, where $\varepsilon(\boldsymbol{k})$ is the electronic dispersion. Crucially, $\mathcal{A}_{ini}$ grows large for small $|\mathbf{G}_{\mathrm{m}}|$, while $\mathcal{A}_{fin}$ is insensitive to $|\mathbf{G}_{\mathrm{m}}|$ in this limit. For small twist angle $\phi$ (small $|\mathbf{G}_{\mathrm{m}}|$) and moderate-energy outgoing electrons, $\mathcal{A}_{ini}/\mathcal{A}_{fin} \gg 1$, the contribution from initial state modification is expected to be the dominant source of moiré replica intensity in the ARPES spectrum.

In situations where the photoemitted electrons originate from a lower monolayer and pass through an upper monolayer, we often observe replicas that are best explained by final-state diffraction from the lattice of the upper layer. For example, in device 3 ($\phi = 9°$) we saw replicas of the WSe$_2$ valence band shifted by reciprocal lattice vectors of the upper WS$_2$ layer (SI Sec. 8). The scattering wave vectors here are long and there is no large parameter that ensures that initial state modification dominates, while the amplitude of the scattering potential can be of atomic scale. A collection of examples of this phenomenon we have seen in 2D heterostructures will be presented elsewhere.

In Figs. 3e-f we see replicas of emission from the capping graphene layer matching the moiré structure of the heterobilayer beneath it. Due again to the small $|\mathbf{G}_{\mathrm{m}}|$ and the fact that the emission is from the topmost layer, these replicas very likely reflect the modification of Bloch states within the graphene layer[44]. The similarity in intensity of the graphene replicas to the parent Dirac cone implies that $\mathcal{A}_{ini}$ here is of the order of unity, indicating that the graphene electrons feel a moiré potential on the order of magnitude of 100 meV. One would expect to see anticrossings on the same energy scale between the replicas and the original bands which would be a clear signature of mini-band formation. The absence of anticrossings in Fig. 3d could be a limitation of the ~100 meV energy resolution.

*6-fold symmetry breaking.* The replicas of both the WS$_2$ and the capping graphene bands in Fig. 3 exhibit an approximate *3-fold* rotational symmetry. Commonly, the moiré superlattice is modeled using a real-valued scalar potential $U(\boldsymbol{r})$, with C$_{3v}$ symmetry. Since $U(\boldsymbol{r})$ is real-valued, and hence its Fourier components satisfy $\widetilde{U}_{-\mathbf{G}_{\mathrm{m}}} = \widetilde{U}^*_{\mathbf{G}_{\mathrm{m}}}$, one might expect the replica intensity pattern to have *6-fold* symmetry. For example, consider a low-energy effective model for the electronic states within one valley, described by the Hamiltonian $H = \hbar v(-i\boldsymbol{\nabla} \cdot \boldsymbol{\sigma}) + \frac{1}{2}\Delta\,\sigma_z + U(\boldsymbol{r})$, where $\boldsymbol{\sigma} = (\sigma_x, \sigma_y)$ and $\sigma_z$ are Pauli matrices representing the orbital pseudospin degree of freedom, $v$ is a velocity and $\Delta$ is a gap. In Fig. 4d we show a schematic representation of $U(\boldsymbol{r})$ in the moiré unit cell. Using the reflection symmetry of the moiré potential across the $y$-axis (the vertical mirror plane $\mathcal{M}_y$ of the C$_{3v}$ point group), $U(x,y) = U(-x,y)$, the model Hamiltonian above is symmetric under the reflection operation $x \to -x$ followed by complex conjugation. As a result, the moiré replicas centered at $\mathbf{G}_1$ and $\mathbf{G}_{-1} = -\mathbf{G}_1$ are represented with equal probability in the modified (perturbed) "initial state" centered around $k = 0$ (i.e., at the valley center). Combined with the 120° rotational symmetry of the system, this would yield a 6-fold symmetric moiré replica pattern.

Crucially, in-plane distortions of the atomic lattice break the mirror symmetry of the system [see, e.g., Ref. [50]]. The resulting local strain fields are manifested in the low-energy effective Hamiltonian through a term of the form $-\hbar v \boldsymbol{A}(\boldsymbol{r}) \cdot \boldsymbol{\sigma}$, where $\boldsymbol{A}(\boldsymbol{r})$ is a moiré (pseudo) vector potential. Physically, this emergent vector potential captures the additional phases acquired by a Bloch wave near the center of one valley as it travels between atomic sites in the strained regions, compared to the phases acquired during hopping in the un-distorted structure. The sign of the moiré pseudo vector potential is *opposite* in valleys **K** and **K'**. This moiré vector potential perturbation breaks the $\mathcal{M}_y$ reflection followed by complex conjugation symmetry of the system that on its own endows the replica intensity pattern with a 6-fold rotation symmetry. The in-plane distortions of the crystal lattice thereby break this 6-fold symmetry down to a 3-fold symmetric pattern. At higher energies where the low-energy effective model is not valid, i.e., sufficiently far from the K point, other factors such as trigonal warping can also give a 3-fold symmetric replica pattern even with only a scalar moiré potential $U(\boldsymbol{r})$. However, the moiré pseudo-vector potential induced 6-fold symmetry breaking persists even close to the K-point. In SI Sec. 9 we analyze this quantitatively within the low-energy continuum model.

**Conclusions**

Underpinning much recent work on correlated and topological states in twisted semiconductor bilayers is the assumption that, far from the zone centre, the bands of the two layers are only weakly hybridized and thus correspond closely to those of the separate monolayers simply superposed. Our results confirm this assumption. In the case of $WS_2/WSe_2$, the band alignment is such that the VB edge is at the K-points in the $WSe_2$ layer, the CB edge is at the K-points in the $WS_2$ layer (with the $WS_2$ Q-point minima just above), and the net band gap is $1.58 \pm 0.03$ eV, all independent of twist angle. In one sample (with 6° twist) we made the first determination of the CB effective mass, finding it to be $0.15 \pm 1\ m_e$ (smaller than predicted). In addition, we observed replicas of the CB shifted in momentum by moiré wavevectors. After theoretically considering the relative contributions of initial-state modification and final-state diffraction, we conclude that the replicas reflect modification of the Bloch states by the moiré potential. The same goes for corresponding replicas of the Dirac cones seen in graphene capping the bilayers. Finally, we consistently observed a three-fold (as opposed to 6-fold) symmetry of the replica pattern which implies that the pseudo-vector potential, and therefore periodic strain, plays a vital role in modifying the Bloch states in moiré structures.

## Methods

**Sample Fabrication**

Standard heated exfoliation and polycarbonate (PC) film-based dry transfer[28] were used. hBN flakes no thicker than 15 nm were chosen to optimize gating efficiency, which under ARPES excitation is negatively impacted by photoexcited carriers in the hBN generating a current to the gate electrode. The Pt/Ti (30nm/5nm) electrodes were predefined using electron-beam lithography. The smaller electrode contacts the graphite gate, as indicated in the optical micrograph in Fig. 1c (see Extended

Data Fig. 1). The larger electrode, which contacts the graphene, is grounded and covers most of the chip to minimize electrostatic distortion of the photoelectron spectrum when applying a gate voltage. The edge of the graphene was patterned into a comb shape because much stronger CB photoemission is obtained from the exposed $WSe_2/WS_2$ heterobilayer between the comb teeth.

All heterostructures are built in three separate parts. First the hBN/graphite back gate is stacked and deposited onto the substrate with electrodes. Then the $WSe_2$ and $WS_2$ are stacked then transferred onto the gate. Finally, the patterned graphene is picked up and transferred onto the rest of the structure. Following each step, the new surface of the substrate or stack is cleaned using contact-mode AFM (Bruker Dimension Icon with OTESPA-R3 cantilevers, setpoint 0.1 V, and line spacing 10 nm/line), with a low velocity (usually 2 μm/s) when cleaning monolayers to minimize tearing. The sample substrates are mounted in dual-inline packages using ultra-high vacuum, high-temperature compatible silver epoxy and wire-bonded. Bare wire is wrapped around the package pins, fixed with epoxy, and are mechanically clamped to leads on the sample mount.

## μARPES with in-situ electrostatic gating

μARPES measurements with in-situ electrostatic gating were conducted at the Spectromicroscopy beamline at Elettra synchrotron[51]. Linearly polarised light was incident on the sample at 45° to the surface normal with a photon energy of 27 eV. The beam was focused to a submicron spot-size on the sample using a Schwarzchild objective. Photoemitted electrons from the top few layers of the sample were collected by an internal moveable hemispherical electron analyser. Under the measurement conditions, the energy and momentum resolution of the 2D detector were roughly 50 meV and 0.03 Å$^{-1}$, respectively. Devices were mounted on sample plates with electrical contacts and annealed at 300 – 350 °C for around 12 hours before measurements. A long anneal time and high temperature were required due to the large size of the chip carrier and sample plate used for in-situ gating. Scanned photoemission microscopy (SPEM) and scanned photocurrent microscopy (SPIM) combined with optical images were used to locate the region of interest on the sample. Energy-momentum slices were obtained by interpolating multiple closely spaced detector images taken along the high symmetry directions of the Brillouin zone. Constant energy maps were extracted from three-dimensional data sets, $I(E, k_x, k_y)$.

## Determining carrier concentrations and $E_F$

The capacitance to the back gate was calculated using $C = \varepsilon_0 \varepsilon_{hBN}/d_{hBN}$, where $\varepsilon_{hBN}$ = 4.5 and the thickness of the hBN was $d_{hBN} = 8.3$ nm (6° sample) or 9.0 nm (2° sample). From the capacitance and electrostatic shift of the electronic bands, $\Delta E_\Gamma$, at each gate voltage $V_g$, the carrier concentration was found from $n_g = C(V_g - \Delta E_\Gamma)/e$. Band shifts, carrier concentrations and photocurrent against gate voltage are displayed in SI Sec. 10 for the 6° twist angle sample. The carrier concentration was then used to find the position of the CB edge, $E_c$, relative to $E_F$ using $n = \int_{E_c}^{\infty} g_c \frac{1}{1 + e^{(E - E_F)/k_B T}} dE$, where $g_c = \frac{g_s g_v m_e^*}{\pi \hbar^2}$ is the density of states of the populated 2D parabolic bands. For the $WS_2$ conduction bands at **K**, the values used were $g_s = 2$, $g_v = 2$, $m_e^* = 0.27$ m$_0$[37] and a splitting of 29 meV[39]. For the $WS_2$ conduction band at **Q**, the values used were $g_s = 1$, $g_v = 6$, $m_e^* = 0.70 m_0$ (average of the two monolayer values[37]) and an energy difference between **Q** and the CBM of 7.1 ± 0.8 meV found from the change in chemical potential between +1.85 V (lowest voltage the conduction band at K is seen in the ARPES spectra) and +3 V (lowest voltage the conduction band at **Q** is seen in the ARPES spectra). For +3 V, $E_F - E_C$ was found to be 2.4 ± 0.5 meV. As $V_g$ is increased, the change in chemical potential is very small when populating the conduction band due to the large density of states and valley degeneracy of the conduction band at **Q**. $E_F$ was found for the gated ARPES spectra by fitting the conduction band at **K** with a Gaussian function multiplied by a sigmoid function; see SI Sec. 10.

**Data availability**

Data presented in this paper are available on request from the authors.

**Author contributions**

NRW, XX and DHC conceived and supervised the project. PVN and HP fabricated the samples. AJG, PVN, JN, VK, MC, AG, NRW and AB collected μ-ARPES data. AJG, and PVN analyzed μ-ARPES data, with input from AB, under the supervision of NRW and DHC. HP acquired photoluminescence data. KW and TT provided the hBN crystals. AA and MR provided theoretical modeling and calculations. DHC, NRW, AJG, HP, PVN and XX wrote the paper with input from all authors.

**Correspondence**

Correspondence and questions for materials should be addressed to cobden@uw.edu, xuxd@uw.edu, or neil.wilson@warwick.ac.uk.

# Supporting Information for:
# Revealing the conduction band and pseudovector potential in 2D moiré semiconductors

Abigail J. Graham[1], Heonjoon Park[2], Paul V. Nguyen[2], James Nunn[1], Viktor Kandyba[3], Mattia Cattelan[3], Alessio Giampietri[3], Alexei Barinov[3], Kenji Watanabe[4], Takashi Taniguchi[5], Anton Andreev[2], Mark Rudner[2], Xiaodong Xu[2,6], Neil R. Wilson[1] & David H. Cobden[2]

[1] *Department of Physics, University of Warwick, Coventry, CV4 7AL, U.K.* [2] *Department of Physics, University of Washington, Seattle, WA, USA.* [3] *Elettra – Sincrotrone Trieste, S.C.p.A, Basovizza (TS), 34149, Italy.* [4] *Research Center for Functional Materials, National Institute for Materials Science, 1-1 Namiki, Tsukuba 305-0044, Japan.* [5] *International Center for Materials Nanoarchitectonics, National Institute for Materials Science, 1-1 Naniki, Tsukuba 305-0044, Japan.* [6] *Department of Materials Science and Engineering, University of Washington, Seattle, WA, USA.*

**Section 1: Piezo-force microscopy phase images for device 1 (6° twisted WS$_2$/WSe$_2$)**

Standard contact-resonant lateral PFM was performed on the heterostructures using a Bruker Dimension Icon AFM with SCM-PIT-V2 cantilevers in a TR probe mount using drive voltage 200-300 mV. Typical contact resonance frequencies were between 650 and 750 kHz. The deflection setpoint is set to saturate the contact resonance amplitude and maintain contact during scanning (usually 0.1V). Coarse topographic mapping for identifying target areas for PFM were done in contact with PFM drive off. The scan surface velocity was kept below 2 μm/s for mapping, and below 0.5 μm/s for PFM measurements.

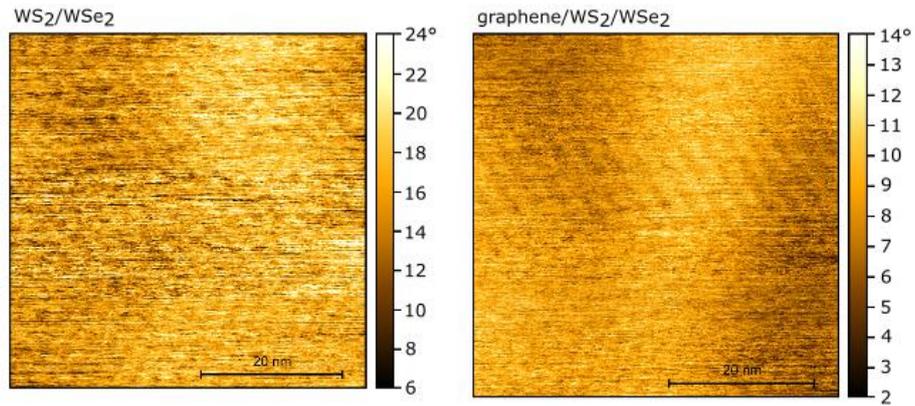

**Figure S1:** Left – PFM phase of exposed WS$_2$/WSe$_2$ region on device 1. Right – PFM phase of graphene-covered WS$_2$/WSe$_2$ region. Scale bars, 20 nm. Both images show period/moiré wavelength of ~2.8 nm, which corresponds to a twist angle of ~6° between the WS$_2$ and WSe$_2$.

**Section 2: Photoluminescence and reflectance measurements to confirm twist angle**

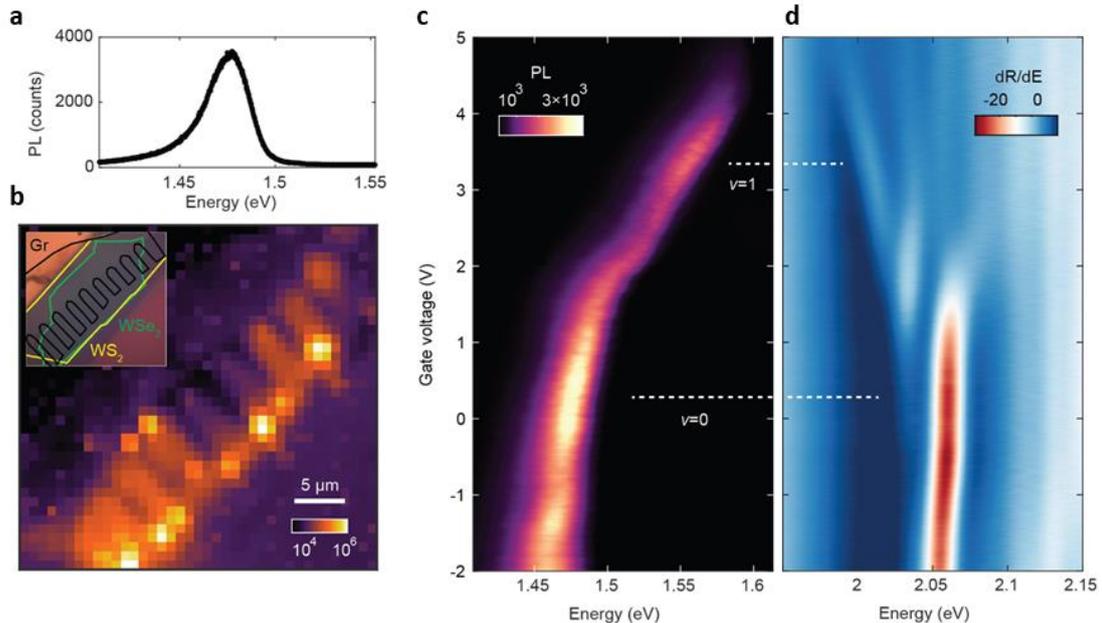

**Figure S3:** a) Interlayer exciton photoluminescence spectrum at charge neutrality measured at 5 K. The sample was excited using a 1.96 eV laser with a power of 10 μW/μm$^2$ and the spectra was integrated for 10 seconds. b) Spatial map of integrated photoluminescence intensity, scale bar 5 μm. The heterobilayer region is relatively bright while the region covered by the graphene comb can be distinguished due to suppression in intensity. c) Gate dependent photoluminescence spectra. d) Gate dependence of energy derivative of reflectance contrast spectrum of the WS$_2$ layer. The filling factor, indicated by white dotted lines, are determined independently through the geometric capacitance, and matches well with the kinks in the two spectra.

### Section 3: Optical, SPEM and SPIM images at multiple gate voltages

Graphene patterned into a comb shape was used as a top contact to the heterobilayers, as shown in Figure S2a. This reduces the contact resistance in the exposed regions of heterobilayer between the graphene teeth of the comb, reducing drop due to in-plane photocurrent generated by the μARPES beam. In Fig. S2b. are scanning photoemission microscopy (SPEM) images which show the integrated photoemission intensity over a range of energy including $E_F$ and a small range in angle near $\mathbf{\Gamma}$. The regions of high integrated intensity reflect the exposed heterobilayer and complement the graphene comb teeth where the integrated intensity is low and uniform. At $V_g$ = 2 V, the high photoemission intensity near $E_F$ persists only near the comb teeth, reflecting where the conduction band is measurably populated. Simultaneously acquired scanning photoemission current microscopy (SPIM) maps, Fig. S2c., give the photocurrent at each point as the beam is rastered, as measured through the graphene top contact. At $V_g$ = 0 the heterobilayer is insulating, so the photocurrent on it is low and the SPIM image shows high photocurrent only on some of the graphene teeth. The absence of photocurrent from other teeth indicates cracks/gaps in the graphene. At $V_g$ = 1.75 V and above, the SPIM images show uniform photocurrent across the heterobilayer in between the graphene teeth, indicating that it is now conductive. Interestingly, the fingers without photocurrent at low voltages appear in SPIM at 1.5 V, suggesting that they are perhaps being bridged to the main graphene body across narrow cracks via the weakly conducting heterobilayer. Analysis of the photocurrent and band position (extracted from the SPEM images) at a region between the graphene teeth is shown in Figure S2d: the energy of the band position varies linearly with applied gate voltage for $V_g$ < 1.5 V, with correspondingly low photocurrent up to this point. For $V_g$ > 1.5 V the photocurrent increases, and the band energy is no longer linearly related to the gate voltage, consistent with the heterobilayer becoming conducting and effectively grounded with a well-defined Fermi energy.

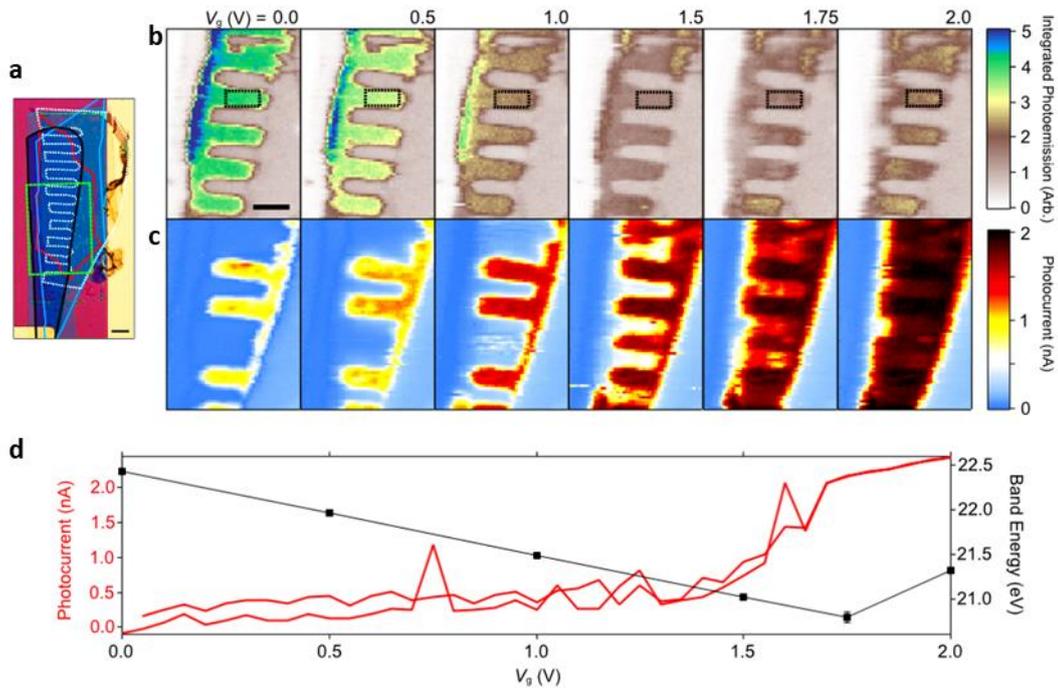

**Figure S2:** a) Optical image of device 1 outlining $WS_2$ (blue), $WSe_2$ (red), cut graphene (dashed white), and graphite (black) regions as transferred. b) SPEM maps (terrain colour scheme) integrated for photoelectron energy $E$ between 20.6 and 24.1 eV with $E_F$ =23.6 eV on a grounded metal electrode and around $\mathbf{\Gamma}$, and c) SPIM maps (cold-hot colour scheme) at multiple gate voltages taken in the area indicated by the yellow dashed box in (a). The bilayer becomes nearly uniformly conducting around 1.75 V. d) Plot of photocurrent (red) and bilayer band position (black) against gate voltage. Band energy found from an average of the black dashed box region in SPEM maps above. All scale bars are 10 μm.

## Section 4: Comparison of monolayer and heterobilayer band alignments and band broadening

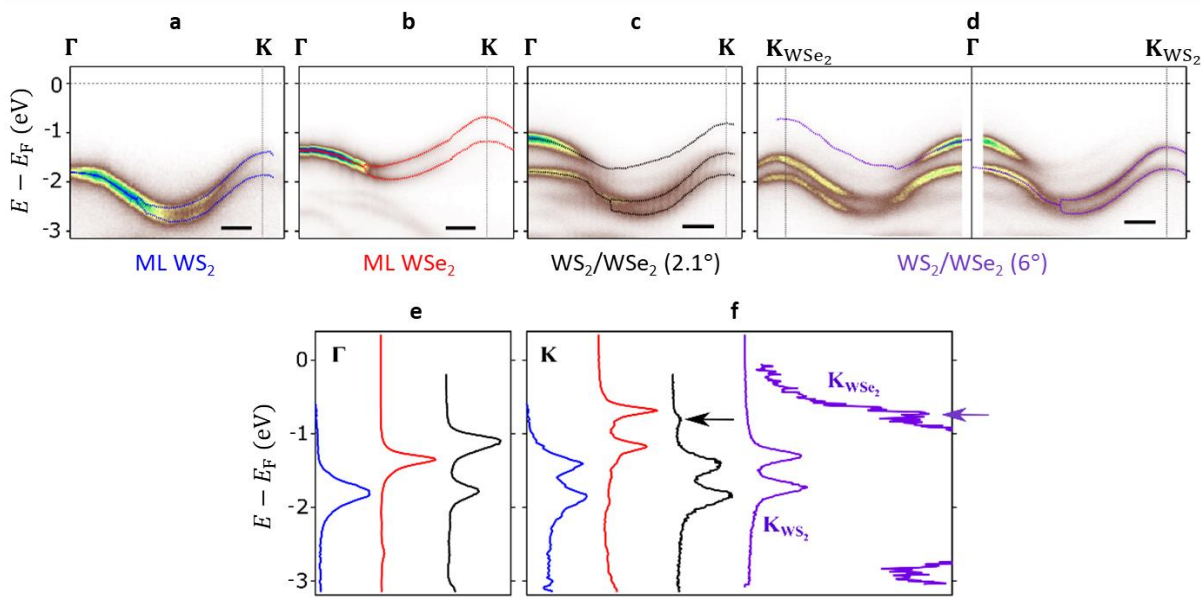

**Figure S4:** a) and b) µARPES energy-momentum slices along the high symmetry direction for WS$_2$ and WSe$_2$ monolayers, respectively. c) and d) µARPES energy-momentum slices along the high symmetry directions defined for the 2° and 6° WS$_2$/WSe$_2$ heterobilayers, respectively. Band fits overlaid. Scale bars, 0.2 Å$^{-1}$. e) and f) Energy distribution curves (EDC) extracted at the high symmetry points, $\Gamma$ and $K$, respectively, from spectra in a-d: ML S$_2$ (blue), ML WSe$_2$ (red), 2° WS$_2$/WSe$_2$ (black), and 6° WS$_2$/WSe$_2$ (purple). Arrows in (f) point to the weak WSe$_2$ band at $K$ for each of the heterobilayers. Band broadenings are typically of order 0.2 eV (full width half maximum). Band widths strongly depend on sample quality and annealing temperatures prior to measurement.

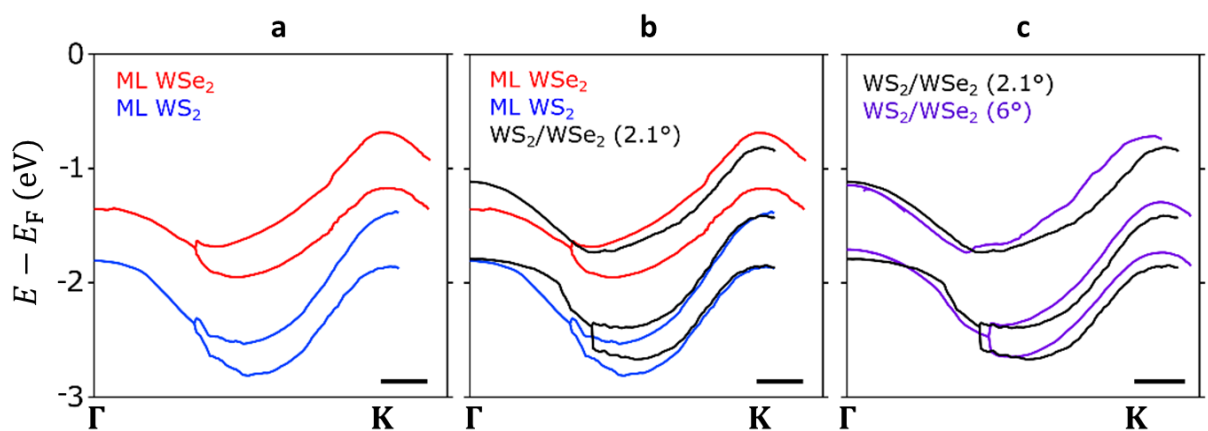

**Figure S5:** a) Fits of the monolayer WS$_2$ (blue) and WSe$_2$ (red) bands from the energy-momentum spectra in Fig. S4. b) Monolayer fits from a) compared to 2° WS$_2$/WSe$_2$ heterobilayer band fits (black). WS$_2$ band structure appears nearly unchanged compared to the monolayer with only a flattening of the band close to $\Gamma$. Upper band at $\Gamma$ is pushed to lower binding due to hybridisation between the layers. c) Comparison of 2.1° and 6° WS$_2$/WSe$_2$ heterobilayer band fits. Small changes in band alignment could be due to a slightly different doping in each sample. Band positions about $\Gamma$ for the 6° heterobilayer are extrapolated due to gap in collected spectra (see Fig. S4d). All fits are also overlaid on the µARPES spectra in Figure S4.

## Section 5: Twist angle dependence of WS$_2$/WSe$_2$ band structure

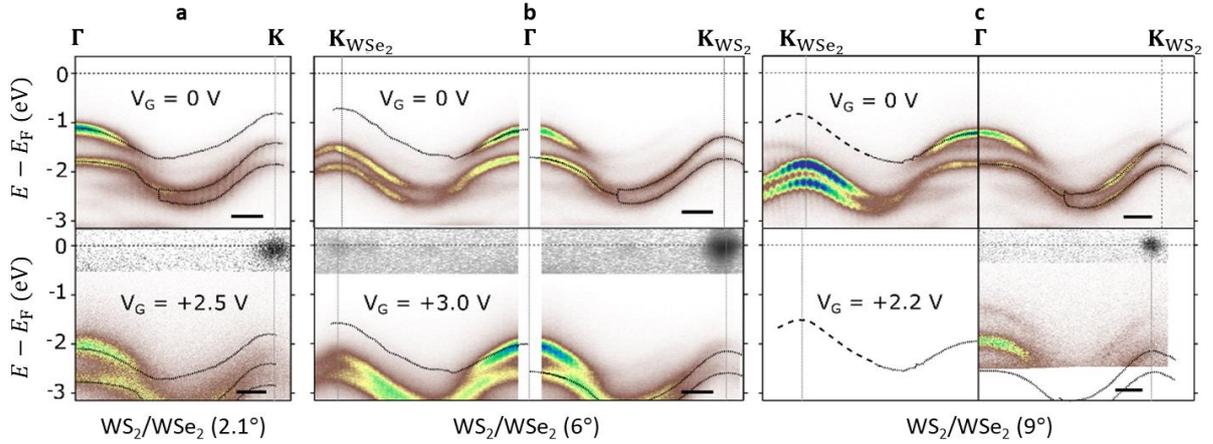

**Figure S6:** WS$_2$/WSe$_2$ band alignments µARPES spectra for three different twist angles: a) 2°, b) 6°, and c) 9°. Dotted lines are fitted band positions. Dashed line is a guide to the eye for the WSe$_2$ uppermost band at **K**.

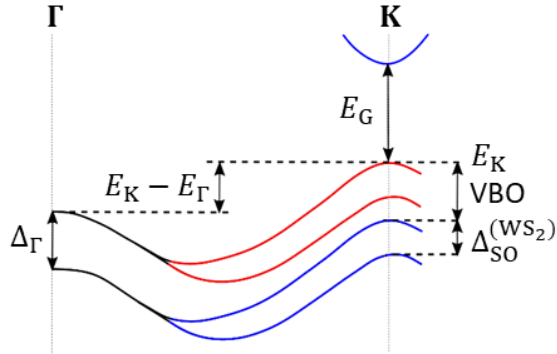

**Figure S7:** Schematic band diagram for WS$_2$/WSe$_2$ heterobilayer. Bands of primarily WSe$_2$ character, WS$_2$ character, and of mixed character due to strong interlayer coupling are red, blue and black respectively. Labelled are the binding energies of the local valence band extrema at **Γ** ($E_\Gamma$) and **K** ($E_K$), the energy differences between the valence band edges at the **Γ**-point ($\Delta_\Gamma$) and at **K** due to spin-orbit coupling in the WS$_2$ ($\Delta_{SO}^{(WS_2)}$), the energy difference between the WSe$_2$ and WS$_2$ valence band edges at **K** (VBO), and the energy difference between the global valence and conduction band extrema ($E_G$).

**Table ST1:** WS$_2$/WSe$_2$ band alignments for each twist angle with notation for band energies tabulated as defined in Fig. S7. $m_K^*$ is the effective mass at the valence band edge at **K** for WS$_2$ / WSe$_2$ respectively. *Uses the energy of the VBM found from the 0 V fit shifted by the electrostatic potential, $\Delta E_\Gamma$.

*[1] for a carrier concentration of $4.2 \times 10^{12}$ cm$^{-2}$. *[2] for a carrier concentration of $6.4 \times 10^{12}$ cm$^{-2}$. *[3] for a carrier concentration of $2.4 \times 10^{12}$ cm$^{-2}$.

| Twist Angle (°) | $E_K$ (eV) | $E_\Gamma$ - $E_K$ (eV) | $\Delta_{SO}^{(WS2)}$ (eV) | VBO (eV) | $\Delta_\Gamma$ (eV) | $m_K^*$ ($m_0$) | $E_G$ (eV) |
|---|---|---|---|---|---|---|---|
| 2 | -0.81 ± 0.03 | -0.31 ± 0.04 | 0.44 ± 0.04 | 0.60 ± 0.04 | 0.67 ± 0.04 | 0.43 ± 0.03 / 0.57 ± 0.05 | 1.62 ± 0.07*[1] |
| 6 | -0.72 ± 0.03 | -0.43 ± 0.04 | 0.44 ± 0.04 | 0.58 ± 0.04 | 0.56 ± 0.04 | 0.38 ± 0.01 / 0.47 ± 0.02 | 1.58 ± 0.03*[2] |
| 9 | -0.92 ± 0.03 | -0.30 ± 0.04 | 0.43 ± 0.04 | 0.52 ± 0.04 | 0.61 ± 0.04 | 0.4 ± 0.1 / not resolved | 1.58 ± 0.06*[3] |

## Section 6: µARPES of WSe$_2$ on WS$_2$.

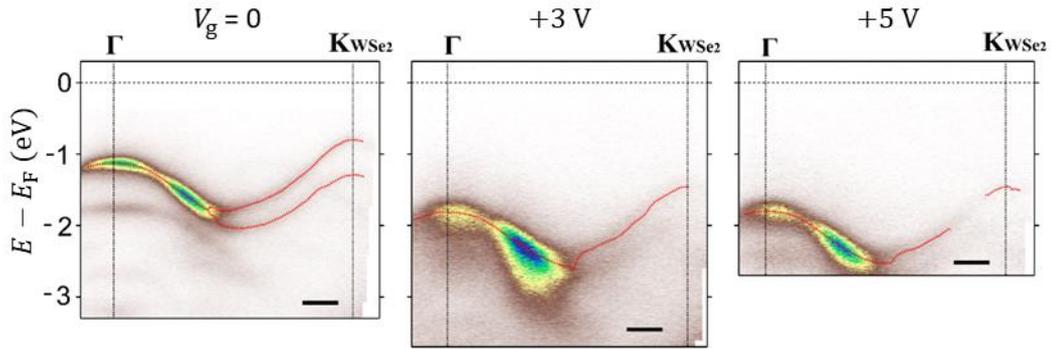

**Figure S8:** µARPES energy-momentum slices from WSe$_2$ on WS$_2$ at gate voltages as shown. The red lines are fits to the upper valence band dispersion, used to find the valence band maximum.

## Section 7: Graphene Dirac cone replicas for WS$_2$/MoSe$_2$ heterobilayer

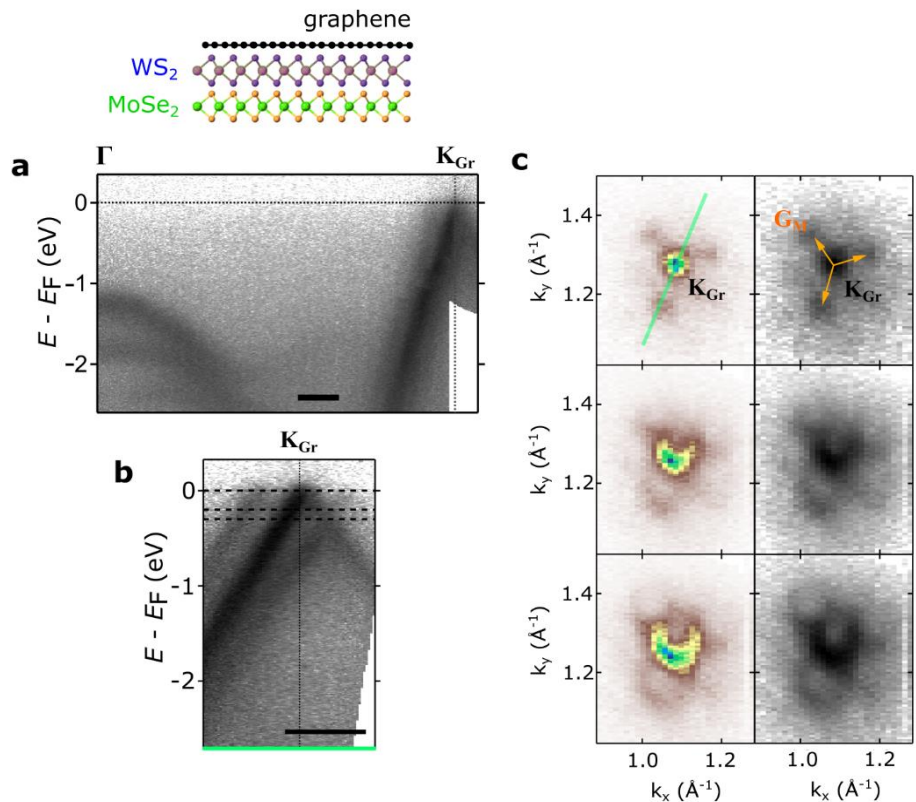

**Figure S9: Moiré-replicas in graphene encapsulated WS$_2$/MoSe$_2$.** (a) Energy-momentum slice along $\Gamma$-$\mathbf{K_{Gr}}$. Above is schematic of the heterostructure. (b) Higher resolution energy-momentum slice around the graphene Dirac cone, showing two moiré replicas either side of the primary band. Scale bars, 0.2 Å$^{-1}$. (c) Constant energy maps taken at the energies of the black dashed lines in (b). Green line in top map in (c) shows the position of the energy-momentum slice displayed in (b). Black and white spectra have normalised photoemission intensity displayed in a log scale. Terrain colour scheme shows normalised photoemission intensity in linear scale. Moiré vector, $G_M$, connecting $\mathbf{K_{Gr}}$ and centres of moiré replicas overlaid in orange.

**Section 8: WS$_2$/WSe$_2$ heterobilayer with 9° twist angle**

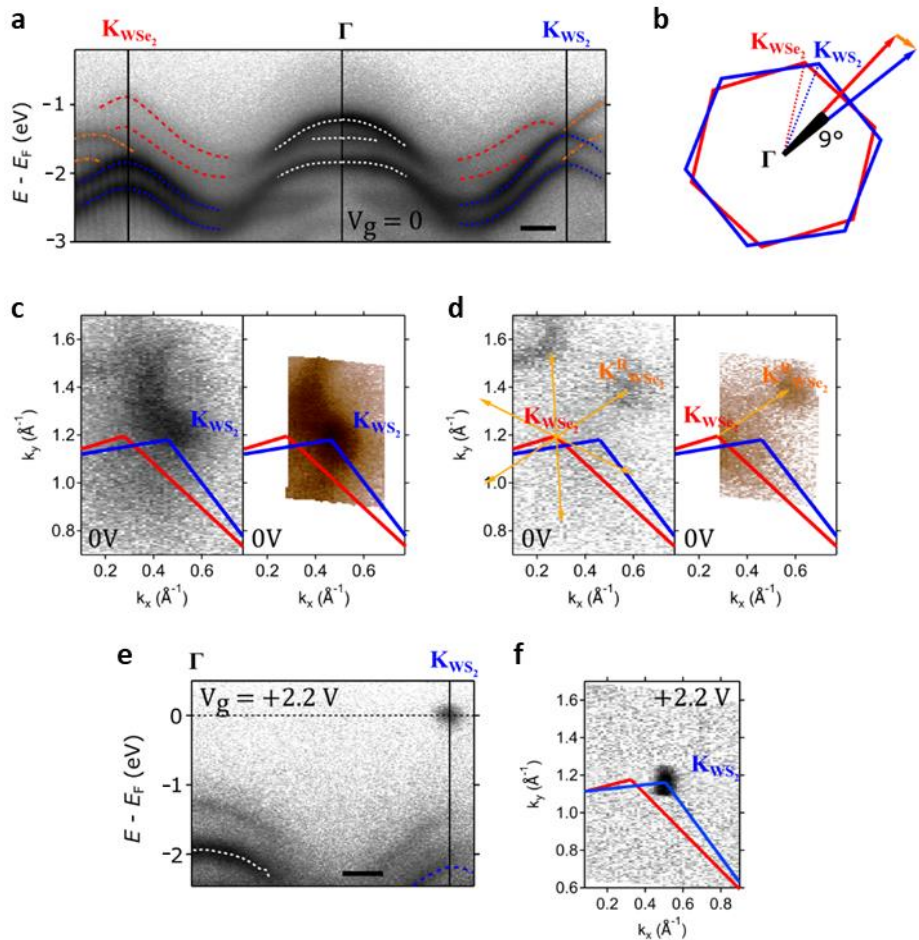

**Figure S10: Photoelectron diffraction replica band in WS$_2$/WSe$_2$ heterobilayer with 9° twist angle.** (a) Energy-momentum slice along the high symmetry directions defined. Overlaid dotted lines are fits of the bands from the WS$_2$ layer. Red and orange dashed lines are lines to guide the eye for the bands from the WSe$_2$ layer and a photoelectron diffracted WSe$_2$ band, respectively. Photoemission intensity displayed in log scale. (b) Schematic of the relative orientation of WS$_2$ and WSe$_2$ Brillouin zones. (c) and (d) Constant energy maps at the energy of the top of the WS$_2$ and WSe$_2$ valence bands, respectively. Map to the right in orange-brown colour scheme is a smaller area higher resolution energy map of the map on the left. Red and blue lines mark the edge of the first Brillouin zone of the WSe$_2$ and WS$_2$ layers, respectively. (e) Energy-momentum slice along the high symmetry direction defined at gate voltage of 2.2 V. White and blue dashed lines correspond with bands in (a). (f) Constant energy map at the Fermi energy at a gate voltage of 2.2 V. Showing no conduction band replicas around the WS$_2$ conduction band at $K$.

## Section 9: Perturbation theory for the modification of the Bloch states by the moiré potential

In this section, we summarize a calculation for the modification of Bloch states by the moiré potential within first order perturbation theory. We particularly focus on the breaking of six-fold symmetry down to three-fold symmetry in the moiré replica intensities. Due to the fact that the three-fold replica contrast is most prominently displayed in the graphene ARPES spectra in the main text, here we focus on the modifications of the Bloch states in graphene by the moiré scalar and vector potentials. The calculation can be straightforwardly extended to describe the TMD layers.

Consider a single sheet of graphene, in the presence of a $C_3$ symmetric moiré potential with scalar and vector contributions $U(\mathbf{r}) = \sum_{\mathbf{G}} U_{\mathbf{G}} e^{i\mathbf{G}\cdot\mathbf{r}}$ and $\mathbf{A}(\mathbf{r}) = \sum_{\mathbf{G}} \mathbf{A}_{\mathbf{G}} e^{i\mathbf{G}\cdot\mathbf{r}}$, respectively, where $\{\mathbf{G}\}$ are the reciprocal lattice vectors of the moiré superlattice (see main text). Due to the fact that the moiré potentials are real-valued, their Fourier components satisfy $U_{-\mathbf{G}} = U_{\mathbf{G}}^*$ and $\mathbf{A}_{-\mathbf{G}} = \mathbf{A}_{\mathbf{G}}^*$. Under a $C_3$ rotation $O$ of the moiré reciprocal lattice vector, $\mathbf{G} \to O\mathbf{G}$, the moiré potential components transform as $U_{O\mathbf{G}} = U_{\mathbf{G}}$ and $\mathbf{A}_{O\mathbf{G}} = O\mathbf{A}_{\mathbf{G}}$. These transformation properties, which follow from the $C_3$ symmetry of the moiré potential, will be important below when we assess the symmetry properties of the moiré replicas in the perturbed wave function.

Focusing on states near valley K, we write the single particle Hamiltonian as $\widehat{H} = \widehat{H}_0 + \widehat{V}$, with $\widehat{H}_0 = v\,\widehat{\mathbf{p}} \cdot \boldsymbol{\sigma}$ and $\widehat{V} = U(\widehat{\mathbf{r}}) + v\mathbf{A}(\widehat{\mathbf{r}}) \cdot \boldsymbol{\sigma}$, where $v$ is the Dirac velocity, $\boldsymbol{\sigma}$ is a vector of Pauli matrices acting on the pseudospin (AB sublattice) space, and we have set $|e| = 1$, where $e < 0$ is the electron charge. For convenience below, we write $\widehat{V} = \sum_{\mathbf{G}} V_{\mathbf{G}} e^{i\mathbf{G}\cdot\widehat{\mathbf{r}}}$, where $V_{\mathbf{G}} = U_{\mathbf{G}} + v\mathbf{A}_{\mathbf{G}} \cdot \boldsymbol{\sigma}$. In the absence of the moiré perturbation, the graphene Bloch band eigenstate in band $n$ (labelling the conduction and valence bands) and at crystal momentum $\mathbf{k}$ (measured relative to the Dirac point) is described by

$$\widehat{H}_0 |\psi_{n\mathbf{k}}^{(0)}\rangle = \varepsilon_{n\mathbf{k}}^{(0)} |\psi_{n\mathbf{k}}^{(0)}\rangle, \quad |\psi_{n\mathbf{k}}^{(0)}\rangle = e^{i\mathbf{k}\cdot\widehat{\mathbf{r}}} |u_{n\mathbf{k}}^{(0)}\rangle,$$

where $|u_{n\mathbf{k}}^{(0)}\rangle$ is the periodic Bloch function and $\varepsilon_{n\mathbf{k}}^{(0)} = \pm\hbar v|\mathbf{k}|$ is the corresponding unperturbed energy, with + and − corresponding to the conduction and valence bands, respectively.

We now seek the perturbed wave function $|\psi_{n\mathbf{k}}\rangle = |\psi_{n\mathbf{k}}^{(0)}\rangle + |\psi_{n\mathbf{k}}^{(1)}\rangle + \cdots$, modified by the presence of the moiré potential. Following a few lines of algebra, the correction to the wave function to first order in the perturbation, denoted $|\psi_{n\mathbf{k}}^{(1)}\rangle$, can be written as

$$|\psi_{n\mathbf{k}}^{(1)}\rangle = \sum_{m\mathbf{G}} |\psi_{m,\mathbf{k}+\mathbf{G}}^{(0)}\rangle \langle\langle u_{m,\mathbf{k}+\mathbf{G}}^{(0)} | \frac{1}{\varepsilon_{n\mathbf{k}}^{(0)} - \widetilde{H}_{\mathbf{k}+\mathbf{G}}} V_{\mathbf{G}} | u_{n\mathbf{k}}^{(0)}\rangle\rangle,$$

where $|u_{n\mathbf{k}}^{(0)}\rangle\rangle$ is the periodic Bloch function restricted to one unit cell, and $\widetilde{H}_{\mathbf{k}}$ is the Bloch Hamiltonian $\widehat{H}_{\mathbf{k}} = e^{-i\mathbf{k}\cdot\widehat{\mathbf{r}}} \widehat{H} e^{i\mathbf{k}\cdot\widehat{\mathbf{r}}}$ restricted to one unit cell. In the two-band tight-binding description, and ignoring spin, $\widetilde{H}_{\mathbf{k}}$ is a $2 \times 2$ matrix acting on the pseudospin degree of freedom.

Using the first-order corrections to the wave functions written above, we define the probability associated with the moiré sideband of the state $|\psi_{n\mathbf{k}}\rangle$ at crystal momentum $\mathbf{k} + \mathbf{G}$, summed over the conduction and valence bands, as

$$P_{\mathbf{k}+\mathbf{G}} = \sum_m |\langle\langle u_{m,\mathbf{k}+\mathbf{G}}^{(0)} | \frac{1}{\varepsilon_{n\mathbf{k}}^{(0)} - \widetilde{H}_{\mathbf{k}+\mathbf{G}}} V_{\mathbf{G}} | u_{n,\mathbf{k}}^{(0)}\rangle\rangle|^2 = \langle\langle u_{n\mathbf{k}}^{(0)} | V_{\mathbf{G}}^{\dagger} \frac{\varepsilon_{n\mathbf{k}}^{(0)} + \hbar v(\mathbf{k}+\mathbf{G})\cdot\boldsymbol{\sigma}}{\left(\left[\varepsilon_{n\mathbf{k}}^{(0)}\right]^2 - |\hbar v(\mathbf{k}+\mathbf{G})|^2\right)^2} V_{\mathbf{G}} | u_{n\mathbf{k}}^{(0)}\rangle\rangle.$$

To evaluate the moiré replica intensity $P_{\mathbf{k}+\mathbf{G}}$ we use $V_{\mathbf{G}} = U_{\mathbf{G}} + v\mathbf{A}_{\mathbf{G}} \cdot \boldsymbol{\sigma}$, along with the identity $(\mathbf{a} \cdot \boldsymbol{\sigma})(\mathbf{b} \cdot \boldsymbol{\sigma}) = \mathbf{a} \cdot \mathbf{b} + i(\mathbf{a} \times \mathbf{b}) \cdot \boldsymbol{\sigma}$. We furthermore use the fact that the pseudospins associated with the graphene Bloch band eigenstates are oriented within the equatorial plane of the Bloch sphere: $\left|\langle u_{n\mathbf{k}}^{(0)} | (\mathbf{z}\cdot\boldsymbol{\sigma}) | u_{n\mathbf{k}}^{(0)}\rangle\right| = 0$. Working in the regime $|\mathbf{k}| \ll |\mathbf{G}|$, after several lines of algebra we obtain the leading contributions in $|\mathbf{k}|/|\mathbf{G}|$:

$$P_{\mathbf{k}+\mathbf{G}} = \frac{|\hbar v \mathbf{G}|^2 - 2(\hbar v)^2 \mathbf{k}\cdot\mathbf{G}}{|\hbar v \mathbf{G}|^4}[U_{\mathbf{G}}^* U_{\mathbf{G}} + v^2 \mathbf{A}_{\mathbf{G}}^* \cdot \mathbf{A}_{\mathbf{G}} + v(U_{\mathbf{G}}^* \mathbf{A}_{\mathbf{G}} + U_{\mathbf{G}} \mathbf{A}_{\mathbf{G}}^*) \cdot \langle\langle\boldsymbol{\sigma}\rangle\rangle]$$

$$+ \frac{2\hbar v\, \varepsilon_{n\mathbf{k}}^{(0)}}{|\hbar v \mathbf{G}|^4}\{vU_{\mathbf{G}}^*(\mathbf{G}\cdot\mathbf{A}_{\mathbf{G}}) + vU_{\mathbf{G}}(\mathbf{G}\cdot\mathbf{A}_{\mathbf{G}}^*) + [v^2(\mathbf{G}\cdot\mathbf{A}_{\mathbf{G}})\mathbf{A}_{\mathbf{G}}^* + v^2(\mathbf{G}\cdot\mathbf{A}_{\mathbf{G}}^*)\mathbf{A}_{\mathbf{G}} + (U_{\mathbf{G}}^* U_{\mathbf{G}} - v^2 \mathbf{A}_{\mathbf{G}}^* \cdot \mathbf{A}_{\mathbf{G}})\mathbf{G}] \cdot \langle\langle\boldsymbol{\sigma}\rangle\rangle\},$$

where we have introduced the shorthand $\langle\langle\boldsymbol{\sigma}\rangle\rangle = \langle\langle u_{n\boldsymbol{k}}^{(0)}|\boldsymbol{\sigma}|u_{n\boldsymbol{k}}^{(0)}\rangle\rangle$. Importantly, in the unperturbed eigenstates, the pseudo spin expectation value is oriented along the direction of $\boldsymbol{k}$: $\langle\langle\boldsymbol{\sigma}\rangle\rangle \propto \boldsymbol{k}$.

According to the symmetry properties outlined above, all terms in the expression for $P_{\boldsymbol{k}+\boldsymbol{G}}$ are invariant under simultaneous 120° rotations of $\boldsymbol{k}$ and $\boldsymbol{G}$, as required by $C_3$ symmetry. To characterize the relative contrast of moiré replicas shifted by the moiré reciprocal lattice vectors $\boldsymbol{G}_{\pm 1, \pm 2, \pm 3}$, we average $P_{\boldsymbol{k}+\boldsymbol{G}}$ over the direction of $\boldsymbol{k}$, for fixed $|\boldsymbol{k}| = k_F$ at the unperturbed Fermi surface: $\bar{P}_{\boldsymbol{G}} = \oint d\theta_{\boldsymbol{k}}\, P_{\boldsymbol{k}+\boldsymbol{G}}$, with $\boldsymbol{k} = k_F(\cos\theta_{\boldsymbol{k}}, \sin\theta_{\boldsymbol{k}})$. After this averaging we obtain

$$\bar{P}_{\boldsymbol{G}} = \frac{1}{|\hbar v \boldsymbol{G}|^2}[U_{\boldsymbol{G}}^* U_{\boldsymbol{G}} + v^2 \boldsymbol{A}_{\boldsymbol{G}}^* \cdot \boldsymbol{A}_{\boldsymbol{G}}] + \frac{\hbar v k_F}{|\hbar v \boldsymbol{G}|^4}(\hbar v \boldsymbol{G}) \cdot v(U_{\boldsymbol{G}}^* \boldsymbol{A}_{\boldsymbol{G}} + U_{\boldsymbol{G}} \boldsymbol{A}_{\boldsymbol{G}}^*).$$

The first term above is invariant under $\boldsymbol{G} \to -\boldsymbol{G}$, and thus gives a six-fold symmetric contribution to the replica intensity. The second term is *odd* under $\boldsymbol{G} \to -\boldsymbol{G}$. Thus we see that while $\bar{P}_{\boldsymbol{G}}$ is invariant under 120° rotations of $\boldsymbol{G}$, i.e., $\bar{P}_{\boldsymbol{G}_1} = \bar{P}_{\boldsymbol{G}_2} = \bar{P}_{\boldsymbol{G}_3}$ (see main text for definitions of $\boldsymbol{G}_{1,2,3}$, the replica probabilities are *not* invariant under 180° (equivalently, 60°) rotations: $\bar{P}_{\boldsymbol{G}_1} \neq \bar{P}_{\boldsymbol{G}_{-1}}$, etc.

**Section 10: Determining band shifts, carrier concentrations, photocurrent, and the conduction band minimum**

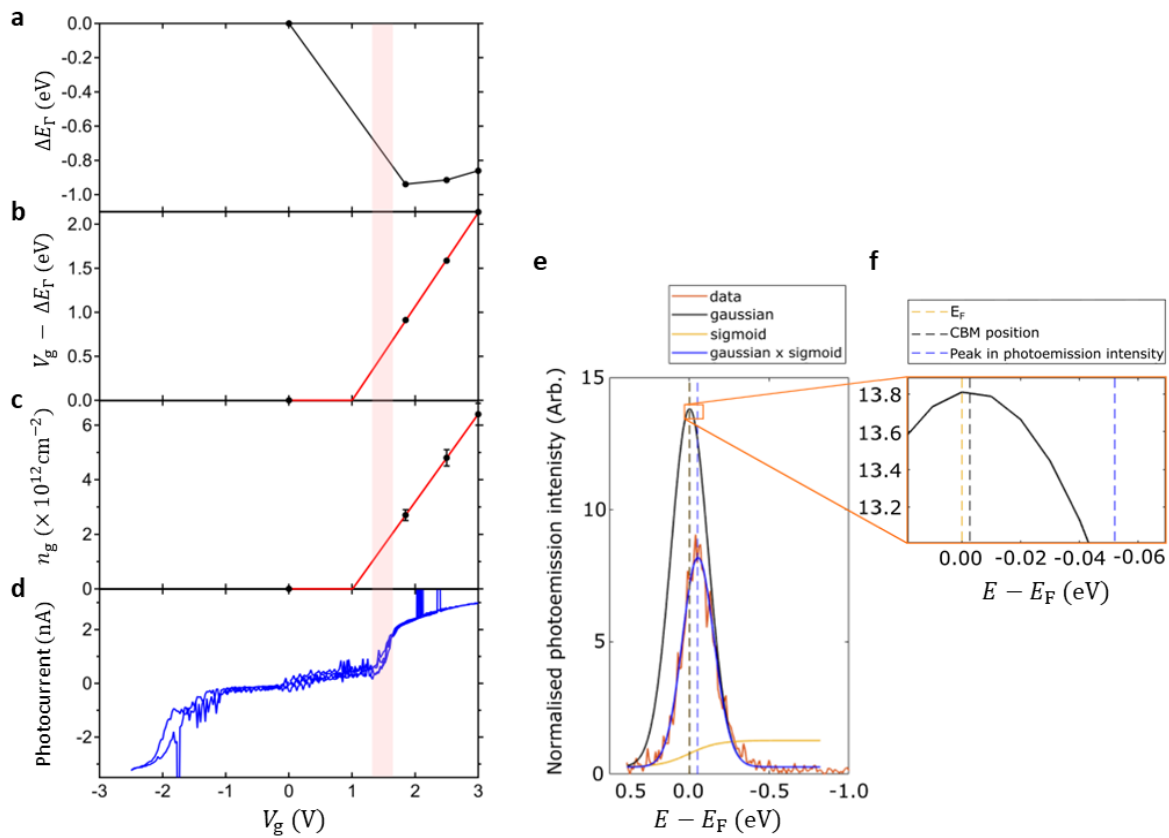

**Figure S11: Gate induced band shifts and photocurrent in WS$_2$/WSe$_2$ heterobilayer with 6° twist angle.** Gate voltage $V_g$ dependence of a) Electrostatic shift of the bands, $\Delta E_\Gamma$; b) Effective potential, $V_G - \Delta E_\Gamma$; c) Carrier concentration, $n_g$; and d) Photocurrent from gate to ground electrode. Pink strip indicates the region where the heterobilayer becomes conducting. e) CBM fit for central EDC shown in Figure 1h. f) zoomed-in region highlighted in orange in e) shows the positions in binding of the CBM and the peak in photoemission intensity relative to the Fermi energy, $E_F$. Peak in photoemission intensity, $E-E_F$ = -0.052 ± 0.003 eV. CBM position = -0.006 ± 0.003 eV. The same Sigmoid function was used in all the EDC fits shown in Figure 1h.